\pdfoutput=1

\documentclass[letterpaper,12pt]{article}

\usepackage{arxiv}
\usepackage[T1]{fontenc}    % use 8-bit T1 fonts
\usepackage{hyperref}       % hyperlinks
\usepackage{url}            % simple URL typesetting
\usepackage{booktabs}       % professional-quality tables
\usepackage{amsfonts}       % blackboard math symbols
\usepackage{nicefrac}       % compact symbols for 1/2, etc.
\usepackage{microtype}      % microtypography
\usepackage{lipsum}
\usepackage{graphicx}
\usepackage[T1]{fontenc} % if needed
\usepackage{graphicx,color}
\usepackage[dvipsnames]{xcolor}
\usepackage{soul}
\usepackage{url}
\usepackage{subcaption}
\usepackage{amssymb}
\usepackage{float}
\usepackage{pdf14}
\usepackage{tikz}
\usepackage{placeins}
\usepackage{tabularx}
\usepackage{schemata}
\usepackage{adjustbox}
\usepackage{graphicx}
\graphicspath{{img/}} %Setting the graphicspath
\usepackage{bm}
\usepackage{listings}
\usepackage{blindtext}
\usepackage{amsmath}
\usepackage{physics}
\usepackage{comment}

\usepackage{aas_macros}
\usepackage{natbib}

\usepackage{color}

\title{Variational autoencoder for generating realistic $N$-body simulations for dark matter halos}

\author{
  Jazhiel Chac\'on-Lavanderos\\ 
  Centro de Investigaci\'on en Computaci\'on, Instituto Polit\'ecnico Nacional,\\
  07738, Ciudad de M\'exico, M\'exico \\
  Instituto de Ciencias F\'isicas, Universidad Nacional Aut\'onoma de M\'exico,\\
  62210, Cuernavaca, Morelos, M\'exico.\\
  \texttt{chaconl2021@cic.ipn.mx} \\
\And
  Isidro G\'omez-Vargas \\
  Department of Astronomy, University of Geneva, \\ 51 Chemin Pegasi, 1290 Versoix, Switzerland.\\
  \texttt{isidro.gomezvargas@unige.ch} \\
\And
 Ricardo Menchaca-Mendez \\
 Centro de Investigaci\'on en Computaci\'on, Instituto Polit\'ecnico Nacional,\\
 07738, Ciudad de M\'exico, M\'exico.\\
 \texttt{ric@cic.ipn.mx}\\
 \And
  J. Alberto V\'azquez \\
  Instituto de Ciencias F\'isicas, Universidad Nacional Aut\'onoma de M\'exico,\\
  62210, Cuernavaca, Morelos, M\'exico.\\
  \texttt{javazquez@icf.unam.mx} }

\begin{document}
\maketitle
\begin{abstract}
{In this paper, we present a deep-learning approach to generate synthetic cosmological images by training a convolutional variational autoencoder on two-dimensional dark matter density slices projected from $\Lambda$CDM $N$-body simulations. The model learns a compact latent representation that enables accurate reconstructions and fast generation of new synthetic realizations through a single forward pass through the decoder. We validate the generated fields using cosmology-based summary statistics, focusing on the matter power spectrum and related Fourier space diagnostics, and found good agreement with the reference simulation across the range of scales where the maps exhibit good resolution. Thanks to its low inference cost and stable training target, this variational-autoencoder approach provides a lightweight and reproducible basis for generative modeling of large-scale projected structures and can support downstream tasks such as fast simulation generation and data augmentation.}

\end{abstract}


%%%---------------------------------------------------------------%%
\section{Introduction}\label{sect:1}
%%%---------------------------------------------------------------%%
 
In cosmology, $N$-body simulations are a fundamental tool to understand the underlying physics of large-scale structure formation. The numerical simulation environment can be considered as a virtual laboratory to rigorously test and validate cosmological models. These simulations typically represent a three-dimensional distribution of millions to billions of particles within a specified volume, evolving under the influence of gravitational and other physical interactions. However, traditional numerical simulations are computationally intensive, as they require repeated calculations to update particle positions and velocities at discrete time steps. Until recently, computational demands limited full-scale cosmological simulations to only a few research groups with extensive resources.

Numerical simulations usually start with initial conditions drawn from a Gaussian random field, evolving particles through gravity (\cite{springel2005cosmological, SvetlinTassev_2013}), dark energy, and additional physical processes such as magnetohydrodynamics (\cite{springel2021simulating, bryan2014enzo}). At each timestep, these simulations solve Poisson's equation numerically, capturing complex, non-linear dynamics and forming non-Gaussian structures such as halos, filaments, and voids. This iterative computational approach inherently leads to substantial computational overhead, highlighting a critical need for more efficient methodologies.

Recently, generative deep learning models have significantly advanced the field of artificial intelligence (AI), offering new possibilities for efficient data generation. Techniques such as Generative Adversarial Networks (GANs) (\cite{goodfellow2016deep, goodfellow2014generative}), diffusion models (\cite{song2020score, chang2023design}) and autoencoders (\cite{KRAMER1992313, NIPS1993_9e3cfc48}) have shown considerable success in generating high-quality synthetic data by learning from training datasets.

Using deep learning, it is now feasible to efficiently emulate the outputs of computationally expensive cosmological simulations. Generative models offer an attractive alternative, significantly reducing computational time while preserving the accuracy and statistical properties crucial for analyzing and understanding cosmological phenomena. Previous works have demonstrated the potential of machine learning in cosmology, such as analyzing dark matter halo structures (\cite{chacon2023analysis, lucie2024deep, bernardini2020predicting, 2021ApJ...915...71V}), and for accelerating numerical processes in cosmology (\cite{gomez2024deep, chardin2019deep, mathuriya2018cosmoflow, to2023linna}).

The generation of 3D cosmological dark matter distributions using techniques such as Generative Adversarial Networks and Diffusion Models presents several challenges due to the complexity and high-dimensional nature of cosmological structures. Unlike simple images, which often contain well-defined objects with clear edges, the large-scale structure of the Universe consists of a continuous field of matter density fluctuations spanning multiple orders of magnitude in scale and density. This makes it difficult for generative models to capture both the fine details of small-scale filaments and the broader statistical properties of large-scale cosmic structures. 

Additionally, training GANs on 3D data can be computationally expensive and less stable in practice \cite{perraudin2019cosmological, 2021A&A...651A..46U, doi:10.1073/pnas.2022038118}, as volumetric data sets require significantly more memory and processing power than their 2D counterparts. Diffusion models \cite{rouhiainen2024superresolution, mudur2022can, mudur2023cosmological, schanz2023stochastic}, while offering advantages in stability and sample diversity, require a large number of inference steps, which makes them computationally prohibitive for large-scale cosmological simulations. Moreover, ensuring that the generated 3D maps preserve key physical properties, such as the matter power spectrum, the halo mass function, and the two-point correlation function, requires careful constraints during training. Often, hybrid approaches incorporate domain knowledge from cosmological simulations. These difficulties highlight the need for specialized architectures and loss functions tailored to the unique statistical and physical constraints of cosmological data. Variational autoencoders, in particular, present promising capabilities to generate realistic cosmological simulations, accurately capturing essential features such as dark matter halos.

Furthermore, when studying large-scale structures in the Universe, the matter power spectrum serves as a fundamental statistical tool that describes the distribution of dark matter as a function of spatial scale, it quantifies the amplitude of density fluctuations at different wavenumbers, offering insights into the growth of structure over cosmic time. The shape and evolution of the power spectrum are highly sensitive to key cosmological parameters, such as the matter density $\Omega_m$, the normalization of the power spectrum $\sigma_8$, and the nature of dark energy through the equation of state parameter $w$. It is key to estimate these values and also interpret data from cosmological surveys such as the Sloan Digital Sky Survey (SDSS) \cite{2000AJ....120.1579Y}, the Dark Energy Survey (DES) \cite{2018PhRvD..98d3526A}, and the upcoming Euclid \cite{2011arXiv1110.3193L} and Vera C. Rubin Observatory's LSST \cite{2009arXiv0912.0201L}. However, there are several challenges in calculating the spectra, including the non-linear evolution of structures, resolution limitations, and other baryonic processes. Numerical simulations such as Illustris \cite{2014Natur.509..177V}, Millennium \cite{2005Natur.435..629S}, and CAMELS \cite{2021ApJ...915...71V}  are widely employed, offering a controlled environment to model the non-linear dynamics of dark matter. Although these simulations provide great statistical tools to understand the physics of the large-scale structures, there are reported discrepancies between simulated and observed spectra, particularly at small scales, where baryonic effects and model assumptions can introduce systematic biases. These issues highlight the need for improved modeling techniques that balance physical realism with computational efficiency.

Variational autoencoders (VAE), in particular, present promising capabilities to generate realistic cosmological simulations, accurately capturing essential features such as dark matter halos. Compared to GANs, VAEs offer greater training stability and a well-defined probabilistic latent space, allowing easier sampling and interpolation between different states. For example, VAEs have been used to model covariance matrices of cosmological datasets (\cite{gomez2023neural}), in CMB image inpainting (\cite{yi2020cosmo}), lensed quasar discovery (\cite{andika2025accelerating}), and gravitational-wave Bayesian inference (\cite{gabbard2022bayesian}).

{Taken together, these previous works motivate a complementary approach that prioritizes computational efficiency, training stability, and controlled statistical validation, rather than maximizing visual fidelity at the highest possible resolution. In this work, we position our contribution as a \emph{lightweight and reproducible baseline} based on a convolutional variational autoencoder that learns a compact latent representation of projected dark-matter density fields. The performance of themodel is primarily evaluated using cosmology-based summary statistics, with particular emphasis on the matter power spectrum. The novelty of this manuscript lies not in the introduction of a new generative architecture, but in demonstrating that a compact VAE baseline can produce statistically consistent realizations at low computational cost and in providing a simple and well-defined reference implementation for generative modeling of projected dark matter density fields in a regime where three-dimensional methods based on GANs or diffusion are typically computationally demanding.
}

{In this work, we aim to develop a generative deep learning framework to emulate the two-dimensional distribution of dark matter in cosmological simulations and to analyze the corresponding matter power spectrum as a function of scale. We train a variational autoencoder using data from an ENZO simulation, described in Section~\ref{sect:3}, and employ the trained model both to reconstruct input density slices and to generate new realizations by decoding latent vectors drawn from the learned prior distribution. The purpose of this work is to produce statistically consistent synthetic samples together with their associated matter power spectra, and to compare these results with state-of-the-art numerical simulations and theoretical expectations.
}

The structure of this paper is as follows. In Section \ref{sect:2}, we provide a brief overview of the numerical methods used for running cosmological simulations, along with key deep learning concepts relevant to the generation of mock cosmological data. Section \ref{sect:3} describes the data set used for training deep generative models, as well as the architecture and implementation details of the models. In Section \ref{sect:4}, we present and analyze the results, focusing on statistical comparisons. This section highlights the advantages of using generative deep learning models—particularly variational autoencoders—as an efficient alternative to full $N$-body simulations for studying large-scale structure formation and extracting cosmological observables such as the matter power spectrum. Finally, in Section \ref{sect:5}, we discuss the implications of our findings, limitations of the current approach, and potential directions for future research and improvements.

%%%---------------------------------------------------------------%%
\section{Methods}\label{sect:2}
%%%---------------------------------------------------------------%%

%%%---------------------------------------------------------------%%
\subsection{$N$-body simulations in cosmology}
%%%---------------------------------------------------------------%%

The dynamical evolution of a system of $N$ particles under gravitational interactions is fundamental for understanding the formation and distribution of large-scale structures in the Universe, such as galaxies and cosmic filaments. The motion of each particle $i$ with mass $m_i$ is governed by Newton's second law:
\begin{equation}
    m_i \frac{d^2 \mathbf{r}_i}{dt^2} = \sum_{j \neq i}^N \mathbf{F}_{ij},
\end{equation}
where $\mathbf{r}_i$ is the position of the particle and $\mathbf{F}_{ij}$ is the gravitational force exerted by the particle $j$ on the particle $i$, given by Newton's law of gravitation:
\begin{equation}
    \mathbf{F}_{ij} = - \frac{G m_i m_j}{(\Delta\mathbf{r}_{ij}^{2} + \epsilon^{2})^{3/2}} (\mathbf{r}_i - \mathbf{r}_j),
\end{equation}
where $G$ is the gravitational constant, $\Delta\mathbf{r}_{ij} = \mathbf{r}_i - \mathbf{r}_j$ and $\epsilon$ is a gravitational softening parameter to avoid division by zero.

In a cosmological context, the equations of motion are often solved in an expanding Universe, described by the Friedmann-Lema\^itre-Robertson-Walker (FLRW) metric \cite{chacon2020dark,chacon2023analysis}. In comoving coordinates $\mathbf{x}_i = \mathbf{r}_i / a(t)$, where $a(t)$ is the scale factor, the equations of motion become:
\begin{equation}
    \frac{d^2 \mathbf{x}_i}{dt^2} + 2H \frac{d\mathbf{x}_i}{dt} = \sum_{j \neq i}^N \frac{G m_j}{a^3} \frac{\mathbf{x}_i - \mathbf{x}_j}{(\Delta\mathbf{x}_{ij}^{2} + \epsilon^{2})^{3/2}},
\end{equation}
where $H = \dot{a}/a$ is the Hubble parameter.

Numerical simulations of the $N$-body problem employ methods such as the Particle-Mesh (PM) algorithm, Tree codes, and Adaptive Mesh Refinement (AMR) or combinations of methods, such as the Tree-Particle-Mesh (Tree-PM) algorithm, to efficiently compute gravitational interactions and track the evolution of cosmic structures. One popular method is the Barnes-Hut algorithm \cite{1986Natur.324..446B}, efficient for computing gravitational interactions in $N$-body simulations by approximating distant particle groups as single massive bodies. 

%%%---------------------------------------------------------------%%
\subsection{Adaptive Mesh Refinement}
%%%---------------------------------------------------------------%%
 
It is a computational technique used in cosmological simulations to dynamically enhance resolution in regions of interest while conserving computational resources. Instead of using a uniform grid, AMR employs a hierarchical, multilevel grid structure where finer grids are generated recursively in high-density regions.

The governing equations in AMR simulations follow the collisionless Boltzmann equation \cite{Vlasov1967TheVP} coupled with the Poisson equation for the gravitational potential:
\begin{equation}
    \frac{d f}{dt} = \frac{\partial f}{\partial t} + \mathbf{v} \cdot \nabla_{\mathbf{x}} f - \nabla \Phi \cdot \nabla_{\mathbf{v}} f = 0,
\end{equation}
where $f(\mathbf{x}, \mathbf{v}, t)$ is the phase-space distribution function, $\mathbf{v}$ is the velocity, and $\Phi$ is the gravitational potential determined by the Poisson equation:
\begin{equation}
    \nabla^2 \Phi = 4 \pi G \rho,
\end{equation}
where $\rho$ is the mass density.

The AMR algorithm refines the grid structure based on a refinement criterion, such as exceeding a density threshold:
\begin{equation}
    \rho > \rho_{\text{threshold}},
\end{equation}
triggering the subdivision of a coarse grid cell into finer cells. This refinement process follows a hierarchical approach, where new levels of finer grids are adaptively introduced to increase resolution only in dynamically significant regions. Mathematically, the AMR grid hierarchy can be described as a series of nested grids with increasing resolution levels $l$, where the grid spacing $\Delta x_l$ at level $l$ is given by:
\begin{equation}
    \Delta x_l = \frac{\Delta x_0}{2^l},
\end{equation}
where $\Delta x_0$ is the resolution of the base grid. The modified equations of motion in the AMR method for collisionless dark matter obey Newton dynamics in comoving coordinates:

\begin{equation}
    \frac{d \mathbf{x} _{i}}{dt} = \frac{1}{a}\mathbf{v}_i,\
\end{equation}

\begin{equation}
    \frac{d\mathbf{v}_i}{dt} = -\frac{\dot{a}}{a}\mathbf{v}_i - \frac{1}{a}(\mathbf{\nabla}\phi)_i,\label{eqn:4.38}
\end{equation}
where $i$ indicates the calculations of the gravitational acceleration at position $\boldsymbol{x}_i$. The gravitational field is then computed from the solutions to the Poisson's equation in comoving coordinates:

\begin{equation}
    \boldsymbol{\nabla}^{2} \phi = \frac{4 \pi G}{a} (\rho - \Bar{\rho}),\label{eqn:4.39}
\end{equation}
where $\rho$ is the local density of the fluid and $\Bar{\rho}$ is the mean density of the fluid \cite{Norman_1999}. This approach provides high resolution in dense regions like galaxy clusters while maintaining efficiency in low-density areas, making AMR an essential tool for large-scale cosmological simulations.

%%%---------------------------------------------------------------%%
\subsection{Autoencoder}
%%%---------------------------------------------------------------%%

An autoencoder is a type of neural network designed for unsupervised learning that aims to compress input data into a lower-dimensional representation (encoding) and then reconstruct it back to its original form (decoding). It consists of two main parts: an encoder, which transforms the input into a compressed latent space, and a decoder, which reconstructs the input from this latent representation. 

The encoder network performs dimensionality reduction, similar to the Principal Component Analysis method \cite{10.1145/3447755}. Furthermore, the autoencoder is optimized for data reconstruction. A good performance of the autoencoder would capture latent variables and perform the decompression process efficiently. 

The model consists of the encoder function $g$ parameterized by $\phi$ and the decoder function $f$ parameterized by $\theta$. The latent space layer captures a low-dimensional representation of an input $x$, this representation is denoted by $\boldsymbol{z}=g_\phi(\boldsymbol{x})$, and the reconstructed output is indicated by $\hat{\boldsymbol{x}}= f_{\theta}(g_{\phi}(\boldsymbol{x}))$. 

Parameters $(\theta, \phi)$ are jointly optimized to produce a reconstructed data sample that closely matches the original data input $\hat{\boldsymbol{x}} \approx f_{\theta}(g_{\phi}(\boldsymbol{x}))$, effectively learning an identity function. Various metrics can be used to measure the discrepancy between two vectors, including cross-entropy when a sigmoid activation function is applied or mean squared error (MSE) for a simpler approach.

%%%---------------------------------------------------------------%%
\begin{comment}
\subsection{Denoising Autoencoder}
%%%---------------------------------------------------------------%%

A Denoising Autoencoder is a modification of the autoencoder designed to prevent the network from learning the identity function. This issue arises when the autoencoder has a high capacity, allowing it to memorize the input data and produce outputs equal to the input, thereby failing to achieve meaningful representation learning or dimensionality reduction. To address this, denoising autoencoders intentionally corrupt the input data by introducing noise or masking certain input values, encouraging the network to learn robust representations capable of reconstructing the original, uncorrupted data.

Given a task $(\mu_{ref}, d)$, to train a DAE, we have to perform the following optimization problem:
%
\begin{equation}
    \min_{\theta, \phi}L(\theta,\phi) = \mathbb{E}_{x \sim \mu_{X}, T\sim \mu_{T}}[d(x,(D_{\theta }\circ E_{\phi }\circ T)(x))].
\end{equation}
%
Usually, the noise process $T$ is applied only during training and testing, not during downstream use. One of the many uses of noise processes includes adding isotropic Gaussian noise, which is a signal noise that has a probability density function (PDF) equal to that of the normal distribution.
\end{comment}

%%%---------------------------------------------------------------%%
\subsection{Variational Autoencoder}
\noindent
%%%---------------------------------------------------------------%%

A Variational Autoencoder is an extension of the traditional autoencoder that incorporates probabilistic modeling to regularize the latent space \cite{kingmaIntroductionVariationalAutoencoders2019}. Unlike a standard autoencoder, which encodes inputs into a fixed latent vector, the VAE encoder outputs two vectors representing the mean ($\boldsymbol{\mu}$) and variance ($\boldsymbol{\sigma}^2$) of a multivariate Gaussian distribution, typically with a diagonal covariance matrix. A latent vector $\boldsymbol{z}$ is then sampled from this distribution and passed to the decoder, which attempts to reconstruct the original input.

Because the encoder outputs a distribution rather than a deterministic vector, the decoder must be able to reconstruct the input across random variations in the latent code. This encourages the model to learn robust and generalizable representations rather than simply memorizing the training data. However, even with this stochastic framework, the model could potentially overfit by generating sharply peaked distributions that are well separated in latent space. To mitigate this, VAEs include a regularization term—specifically the Kullback–Leibler (KL) divergence \cite{foster2022generative, kullback1951information}—that penalizes deviations from a predefined prior distribution, usually a standard normal distribution $\mathcal{N}(\boldsymbol{0}, \boldsymbol{I})$.

This regularization induces a continuous and smooth latent space that closely aligns with the support of the standard multivariate Gaussian distribution. As a result, similar inputs are mapped to nearby regions in the latent space. Furthermore, the VAE can be used as a generative model: new data samples can be produced by drawing latent vectors from the prior distribution and decoding them into the input space.

The VAE is trained by maximizing the Evidence Lower Bound (ELBO), which serves as a lower bound on the log-likelihood of the observed data \cite{luoUnderstandingDiffusionModels2022}. The ELBO can be expressed as:
\begin{equation}
    \text{ELBO} = \mathbb{E}_{q_{\phi}(\boldsymbol{z}|\boldsymbol{x})}[\log p_{\theta}(\boldsymbol{x}|\boldsymbol{z})] - D_{KL}(q_{\phi}(\boldsymbol{z}|\boldsymbol{x})||p(\boldsymbol{z})),
\end{equation}

where $q_{\phi}(\boldsymbol{z}|\boldsymbol{x})$ is the approximate posterior distribution of the latent variable $\boldsymbol{z}$ given the input $\boldsymbol{x}$, and $p_{\theta}(\boldsymbol{x}|\boldsymbol{z})$ is the likelihood of the input given the latent variable. The first term encourages the model to reconstruct the input accurately, while the second term regularizes the latent space to follow a standard normal distribution.

To enable gradient-based optimization, the reparameterization trick \cite{kingmaIntroductionVariationalAutoencoders2019} is applied, which allows the sampling of latent variables while preserving the differentiability during backpropagation. Specifically, the latent variable $\boldsymbol{z}$ is reparameterized as:
\begin{equation}
    \boldsymbol{z} = \mu + \sigma \odot \epsilon,
\end{equation}

where $\epsilon \sim \mathcal{N}(0, I)$ is a standard normal noise vector and $\odot$ denotes element-wise multiplication. This reformulation allows the stochastic sampling process to be expressed as a deterministic and differentiable function of $\boldsymbol{\mu}$ and $\boldsymbol{\sigma}$, enabling efficient training through standard backpropagation techniques.

To optimize the VAE, the loss function combines two terms: (1) reconstruction loss, which measures the difference between the input and output using the mean squared error, and (2) Kullback-Leibler (KL) divergence, which regularizes the latent space by ensuring that the learned distribution remains close to a standard Gaussian prior. The total loss is defined as:
\begin{equation}
    \mathcal{L} = \mathbb{E}_{q_{\phi}(z|x)}[||x - \hat{x}||^{2}] + D_{KL}(q_{\phi}(z|x)||p(z)).
\end{equation}

%%%---------------------------------------------------------------%%
\section{Data training}\label{sect:3}
\noindent
%%%---------------------------------------------------------------%%
 
We utilized the ENZO Cold Dark Matter (CDM) simulation dataset \cite{bryan2014enzo}, which was generated using a set of well-established cosmological parameters: a dark matter density of $\Omega_{DM} = 0.25$, a baryonic matter density of 
$\Omega_b = 0.05$, a dark energy density of $\Omega_\Lambda = 0.7$ and a Hubble parameter of $h = 0.71$ km/s/Mpc. The simulation spans a comoving cubic volume of 128 Mpc/$h$ per side, evolving from an initial redshift of $z_i = 99$ to a final redshift of $z_f=0$ and a total of $64^{3}$ particles. The computational domain is initialized with a base root grid\footnote{The base root grid provides the initial spatial resolution for modeling physical processes and contains the coarsest representation of the simulated region before adaptive mesh refinement (AMR) techniques enhance resolution where necessary.} of $64^3$
cells, ensuring an appropriate resolution for capturing the underlying structure formation.

To construct our dataset, we selected simulation snapshots spanning the redshift range from $z=2$ to $z=0$. This range encapsulates the late-time evolution of the large-scale structure, allowing the model to learn key features of structure formation. {However, all latent-space sampling, image generation, and quantitative analyses presented in this work are restricted to snapshots at redshift $z=0$. This choice ensures that the generated realizations correspond to a fixed cosmological epoch, facilitating direct statistical comparisons in the fully evolved nonlinear regime.}

The training samples were constructed by extracting two-dimensional slices from the three-dimensional density field of the simulations. Each slice represents a projected view of the dark matter density field along different axes ($x-y$, $x-z$ and $y-z$), capturing the anisotropic nature of cosmic structure formation. This projection technique provides a more comprehensive dataset, ensuring that the model is exposed to a diverse set of density distributions. Additionally, by incorporating random rotations and noisy variations of the images, we further enhance the robustness of the model, enabling it to generalize better across different realizations of the dark matter distribution.

%%%---------------------------------------------------------------%%
\subsection{Variational Autoencoder Architecture and Training}
\noindent
%%%---------------------------------------------------------------%%
 
To generate high-fidelity representations of cosmic structures, we implemented a Variational Autoencoder designed to learn an efficient latent space representation of dark-matter density distributions. The VAE consists of an encoder, a latent space, and a decoder, each optimized to reconstruct cosmological density fields while enforcing a structured latent representation.

The encoder architecture consists of a sequence of convolutional layers with increasing filter depths (64, 128, 256, and 512 channels), which progressively downsample the input and extract hierarchical spatial features. These are followed by two fully connected layers that map the learned representations to the mean $\mu$ and logarithmic variance $\log \sigma^2$ of the latent space distribution. To enhance model performance, we conducted a systematic hyperparameter optimization to determine the optimal kernel size for the convolutional layers, the number of units in the dense layers, and the training batch size.

Training was conducted using the Adam optimizer with an initial learning rate of $10^{-4}$, combined with an early stopping mechanism to prevent overfitting and a learning rate scheduler to adaptively refine convergence. The model was trained for up to 80 epochs, leveraging data augmentation techniques such as random rotations and additive Gaussian noise to improve generalization.

To identify an optimal architectural configuration, we performed a multi-objective hyperparameter search aimed at jointly minimizing the reconstruction loss and the Kullback–Leibler (KL) divergence. Following the suggestion in \cite{gomez2023neuralgenetic}, which demonstrates the suitability of genetic algorithms for complex spaces in cosmological modeling, we adopted an evolutionary optimization strategy. Specifically, we employed the NSGA-II (Non-dominated Sorting Genetic Algorithm II), as implemented in the \texttt{Optuna} optimization framework \cite{akiba2019optuna}. Table \ref{tab:hyperparams} summarizes the hyperparameter space considered and the best values found during the optimization process.

\begin{table}[h]
    \centering
    \caption{Hyperparameters used in the VAE architecture and best configuration found via Optuna with NSGA-II.}
    \label{tab:hyperparams}
    \begin{tabular}{@{}lll@{}}
        \toprule
        \textbf{Hyperparameter} & \textbf{Search Space}        & \textbf{Best Value} \\ \midrule
        Batch size              & \{2, 4, 8\}                  & 4                   \\
        Kernel size             & \{3, 5\}                     & 5                   \\
        Dense units             & \{64, 128, 192, 256\}        & 256                 \\
        % Learning rate           & Fixed at $10^{-4}$           &                    \\
        % Latent dimension        & Fixed at 512                 &                    \\
        % Epochs                  & Early stopping ($\leq$ 50)   &                    \\
        \bottomrule
    \end{tabular}
\end{table}

The decoder follows the structure of the encoder, using transposed convolutional layers to upsample the latent representation and reconstruct the input image. It starts with a dense layer reshaped into a lower resolution feature map, followed by a series of deconvolutional layers that gradually increase the spatial resolution while maintaining the features learned by the encoder. The design of the decoder and encoder is based on the hyperparameters presented in Table~\ref{tab:hyperparams}, which were obtained through the optimization process described in the previous paragraphs. This configuration allows the latent space to capture meaningful physical information while ensuring an accurate reconstruction of the input density fields. The complete implementation of the VAE architecture and the training procedure is publicly available\footnote{\url{https://github.com/igomezv/NcosmoVAE}}.

%%%%%%%%%%

%%%---------------------------------------------------------------%%
\section{Results}\label{sect:4}
\subsection{Statistical Evaluation of the VAE-Generated Dark Matter Distributions}
\noindent
%%%---------------------------------------------------------------%%
 
We evaluated the performance of the Variational Autoencoder by performing a quantitative evaluation of the statistical properties of the generated samples. We focus on comparing numerical descriptors that characterize the spatial distribution of matter across the dataset. This approach allows us to determine whether the VAE captures the underlying statistical structure of the training data in the context of cosmological large-scale structure formation.

\begin{figure}[t]
    \centering
\begin{minipage}[b]{0.49\textwidth}
    \includegraphics[height=6.8cm, width=8cm]{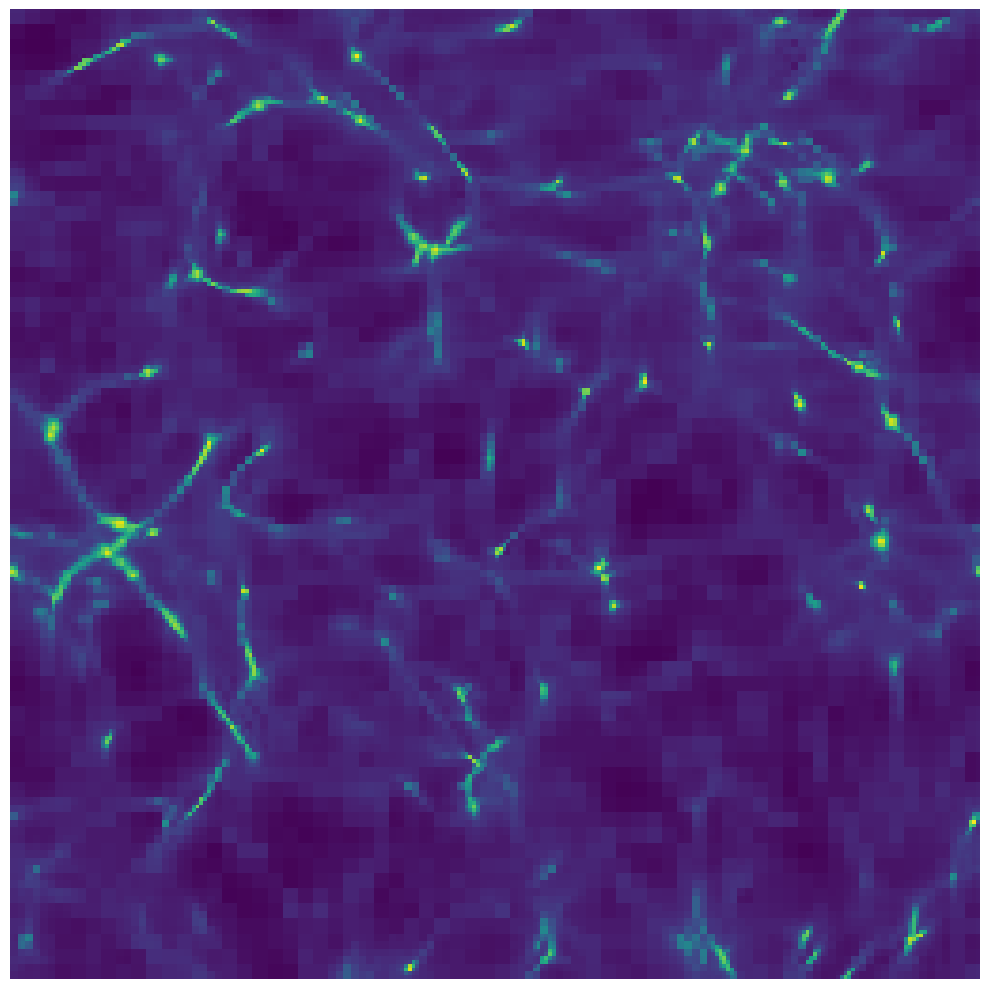}
 \end{minipage}
  \begin{minipage}[b]{0.49\textwidth}
    \includegraphics[height=6.8cm, width=8cm]{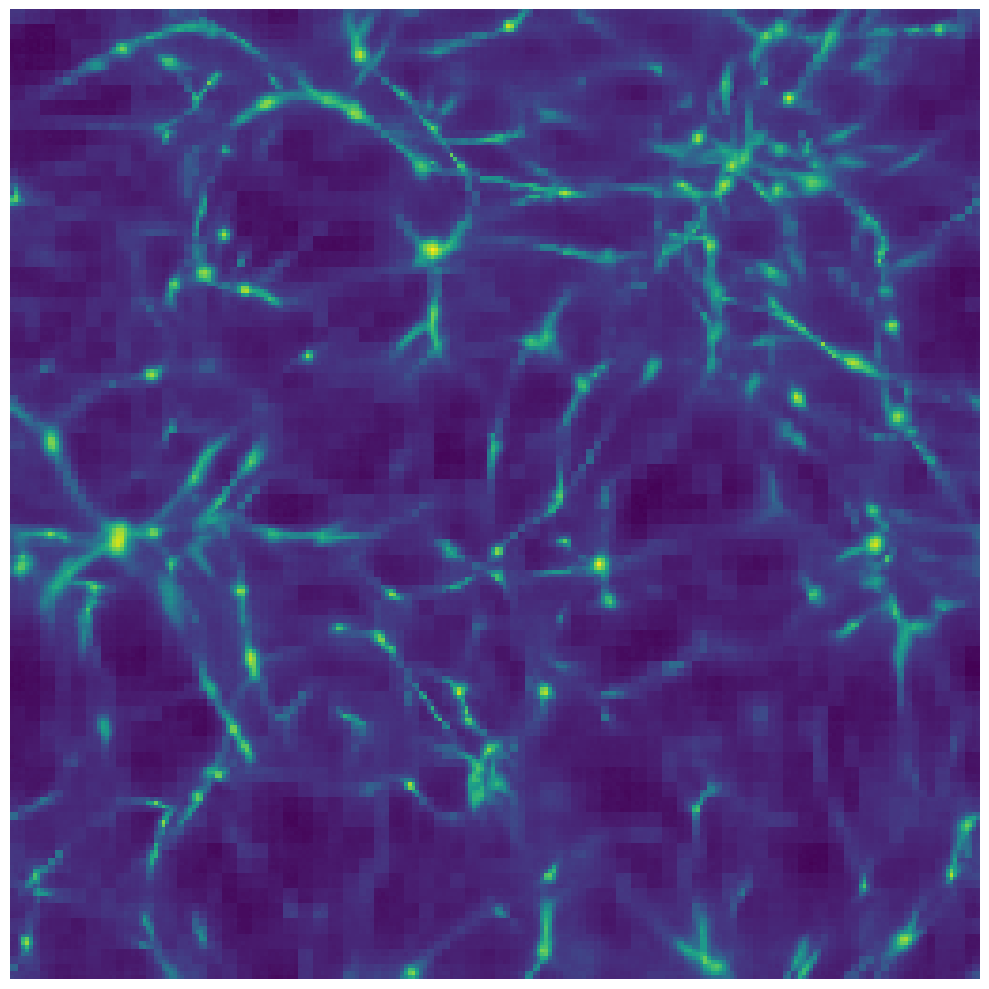}
  \end{minipage}

\caption{Comparison between a dark matter density field from the training dataset (left) and a sample generated by the VAE model (right), shown at identical resolution and dynamic range. While both fields exhibit similar large-scale structure, the VAE-generated sample displays smoother features, reflecting the model’s ability to capture global statistical properties of the density field while regularizing fine-scale variations.}
\label{fig:1}
\end{figure}

To perform this evaluation, we compute summary statistics from both the original and generated density fields, including the power spectrum and pixel intensity histograms. These measures offer insight into the ability of the VAE to reproduce the clustering behavior, density fluctuations, and hierarchical features typical of dark matter distributions. In Figure~\ref{fig:1}, both the training (left) and VAE-generated (right) density fields are shown. The generated samples exhibit a strong resemblance to the training data, indicating that the VAE preserves the overall statistical properties of the distribution. However, the generated fields tend to display smoother features and slightly less fine-grained detail. This behavior reflects the regularizing effect of the VAE and highlights its potential to generate statistically consistent samples for data augmentation purposes.

\begin{figure}[h]
    \centering
    \includegraphics[width=1\linewidth]{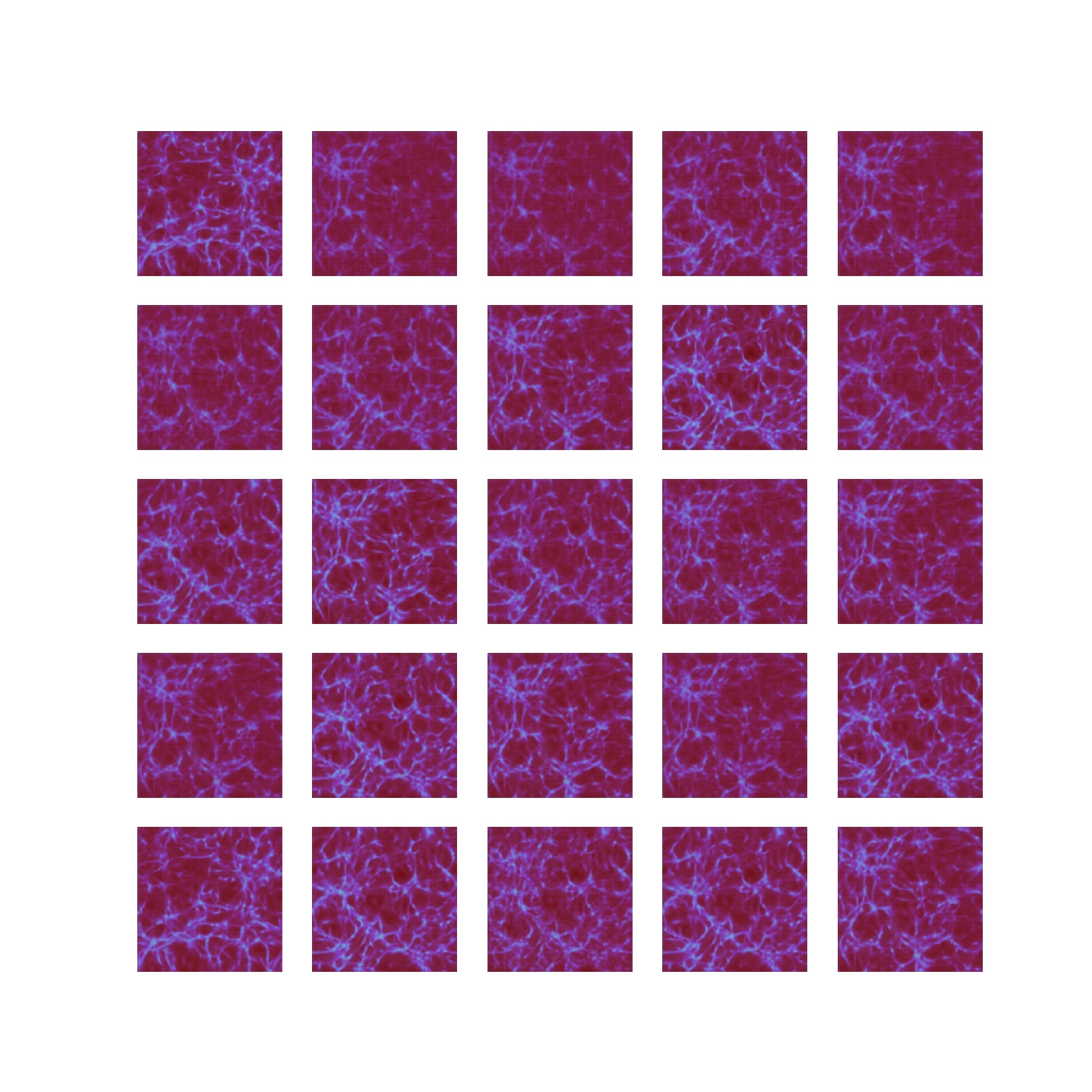}
    \caption{A 5×5 grid of dark matter density fields generated by the VAE using random samples from the latent space. The diversity and consistency of the structures demonstrate that the model learns a meaningful latent representation, enabling the generation of novel, statistically coherent realizations that reflect the large-scale features present in the training data.}
    \label{fig:2}
\end{figure}

Furthermore, when analyzing randomly drawn generated images, we observe that they exhibit structural features that are qualitatively similar to those seen in the training data. Figure~\ref{fig:2} shows a set of samples generated by drawing random vectors from the VAE's latent space after training. The resulting density fields retain the large-scale structural characteristics of the training dataset, such as filamentary patterns and voids, while exhibiting diversity that reflects the stochastic nature of the latent space. {This behavior indicates that the VAE learns a meaningful and generalizable representation of the underlying data distribution rather than memorizing individual training samples. The generated outputs are therefore statistically consistent with the input data, while remaining distinct realizations. As expected for a probabilistic generative model, some loss of fine-scale detail is observed in certain cases, particularly in regions of the latent space that are less densely sampled, reflecting an inherent averaging effect of the VAE formulation.}

{To assess computational efficiency, we compared the runtime of our VAE-based approach with that of an ENZO cosmological simulation. Although a direct comparison of core-hours is not strictly equivalent due to the fundamentally different nature of the methods, it nevertheless provides a useful reference for computational cost. The ENZO simulation was performed on a 16-core CPU architecture and required a total of 720 core-hours. In contrast, the VAE-based pipeline required approximately 3 hours of wall-clock time on a single NVIDIA T4 GPU. This corresponds to a reduction of more than an order of magnitude in computational time, highlighting the potential of deep generative models as scalable tools for cosmological data emulation and exploratory analysis. Beyond training, sample generation with the VAE requires only a forward pass through the network. This contrasts with diffusion-based models, which rely on iterative denoising steps at inference, and with full $N$-body simulations, whose computational cost scales at least as $\mathcal{O}(N \log N)$ with the number of particles. While GANs can achieve comparable inference complexity, their training is typically more unstable and computationally demanding, whereas diffusion-based approaches constitute a promising direction for future work.}

\noindent
The statistical properties of large-scale structures can be quantitatively assessed using the matter power spectrum, which characterizes the distribution of matter in Fourier space. When using a VAE to generate synthetic cosmological images, it is crucial to evaluate how well the generated images preserve the statistical properties of the training dataset. An effective method for this comparison is through power spectral analysis, which provides insights into the spatial correlations and hierarchical clustering present in both real and generated data.

The matter power spectrum is computed by taking the Fourier transform of the density field extracted from the images. Mathematically, it is given by:

\begin{equation}
    P(k) = \bra{} \overset{\sim}{\delta}(k) \ket{^{2}},
\end{equation}
where $\overset{\sim}{\delta}(k)$ is the Fourier transform of the matter overdensity field and represents the wavenumber corresponding to different spatial scales.

To ensure a rigorous statistical assessment, we computed the power spectrum for both the original training images and the generated images. This allows us to analyze whether the VAE effectively captures the underlying statistical distribution of matter in the universe. The comparison is performed by:
\begin{enumerate}
    \item Transforming both the training and the generated images into density fields.

    \item Computing the discrete Fourier transform to obtain the power spectrum.

    \item Averaging over multiple realizations to reduce statistical noise.
 
\end{enumerate}

%\begin{figure}[t]
%    \centering
%\begin{minipage}[b]{0.49\textwidth}
%    \includegraphics[height=6.8cm, width=8cm]{fake_sample.png}
% \end{minipage}
%  \begin{minipage}[b]{0.49\textwidth}
%    \includegraphics[height=6.8cm, width=8cm]{fake_sample.png}
%  \end{minipage}
%\end{figure}

\begin{figure}[t]
    \centering
\begin{minipage}[b]{0.49\textwidth}
    \includegraphics[height=6.8cm, width=8cm]{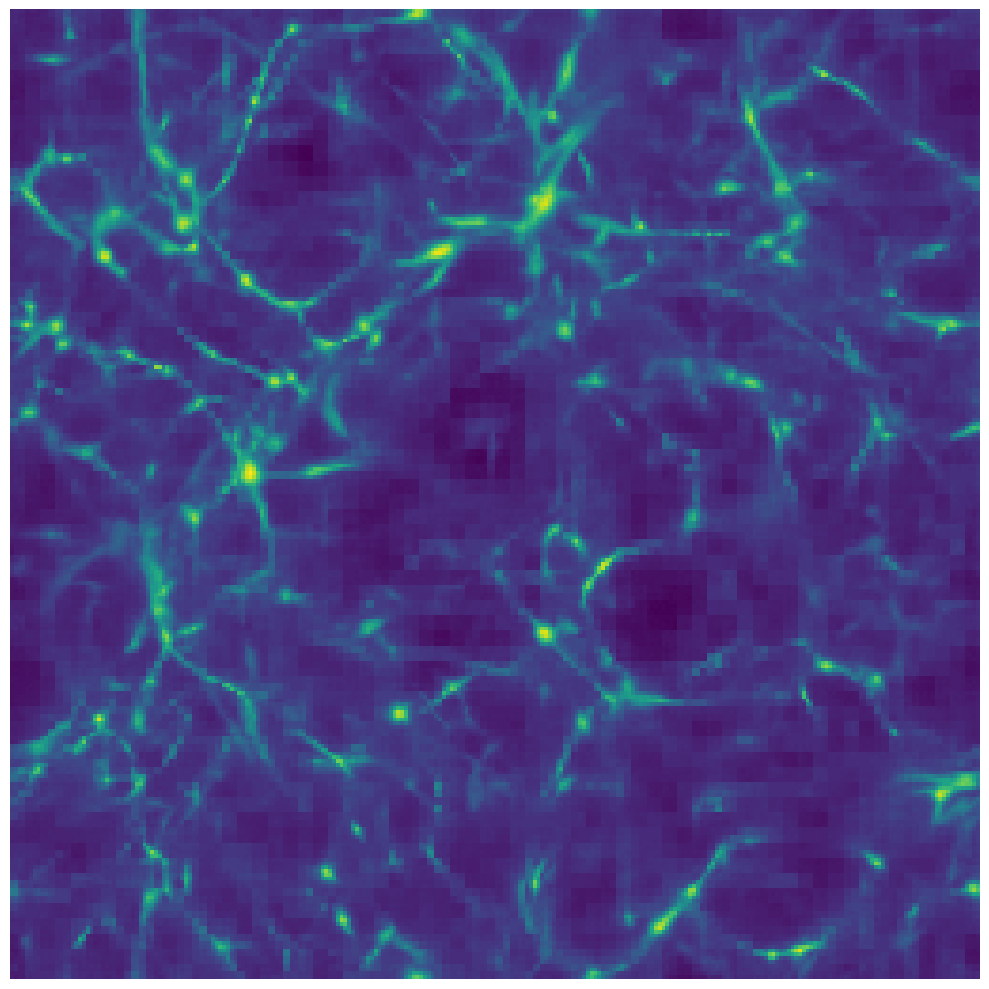}
 \end{minipage}
  \begin{minipage}[b]{0.49\textwidth}
    \includegraphics[height=6.8cm, width=8cm]{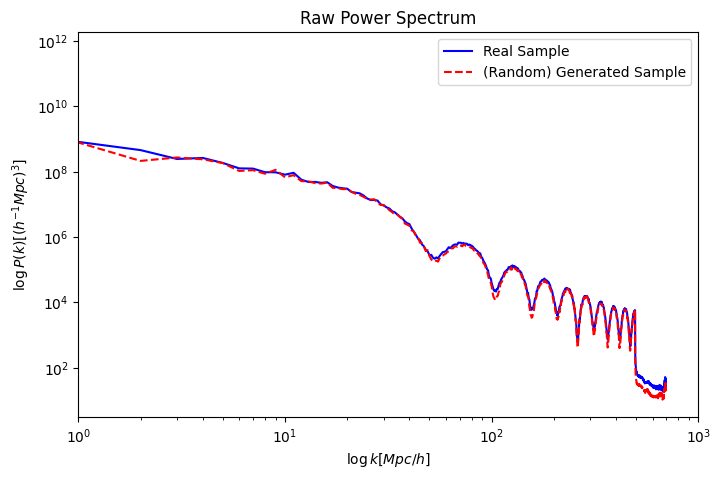}
  \end{minipage}
  \caption{\footnotesize {Raw power spectra for generated dark matter density fields. The top row shows a reconstruction: the left panel displays a VAE reconstruction of a real input slice, and the right panel shows its corresponding power spectrum (red dashed line), compared with the original sample (blue solid line). The bottom row shows an \emph{unconditional} generation from a random latent vector (left), along with its power spectrum (right). In both cases, the VAE reproduces the large-scale (low $k$) behavior of the real data, while deviations at smaller scales (high $k$) highlight the smoothing effects introduced by the generative process.}}

  \label{fig:2.5}
\end{figure}

Preliminary results indicate that the VAE successfully replicates the broad trends in the matter power spectrum, particularly on intermediate spatial scales. However, deviations at small scales (high $k$-values) suggest that fine-grained features, such as filamentary substructures and small halos, may not be fully captured by the model. Such discrepancies could arise due to the inherent smoothing effects introduced by the VAE’s latent space or the limitations of the training dataset. Figure~\ref{fig:2.5} presents two rows of power spectra: the top row corresponds to a reconstructed image generated by the VAE, while the bottom row shows the spectrum of a sample generated from a random latent vector. Both spectra exhibit similar behavior on intermediate scales, indicating the ability of the model to capture the overall statistical structure. However, at high $k$ values (small scales), the characteristic oscillations of the power spectrum are more pronounced, possibly reflecting differences in how the model encodes small-scale information under random sampling.  

The wavenumber scale $k$ was corrected under the assumption that the image represents a comoving box
$128\; \text{Mpc}/h$, allowing spatial frequencies from the Fourier analysis to be converted into physical units of $h/\text{Mpc}$. At large scales (small $k$ values), the power spectrum exhibits the expected smooth decline, consistent with a nearly scale-invariant primordial behavior. Around $k \sim 0.2 \; \text{Mpc}/h$, a transition into the nonlinear regime is observed, marked by a change in the spectral slope. However, small oscillations and artificial peaks appear at high frequencies, likely due to aliasing, discrete sampling, or compression artifacts in the original image, introducing noise into the spectral analysis.

To address these deviations, we implemented a statistical filtering approach to selectively suppress noise-dominated modes at high $k$-values while preserving the physically meaningful large-scale structure. After applying this scale-dependent filter, the agreement between the generated and target power spectra improved significantly, particularly in the transition region between linear and nonlinear scales (0.1 < $k$ < 1.0 Mpc$^{-1}$). This post-processing step demonstrates that while the raw VAE output exhibits limitations at small scales, its predictions can be refined through targeted filtering to better match the expected cosmological statistics.

\begin{figure}[t]
    \centering
    \includegraphics[width=0.5\linewidth]{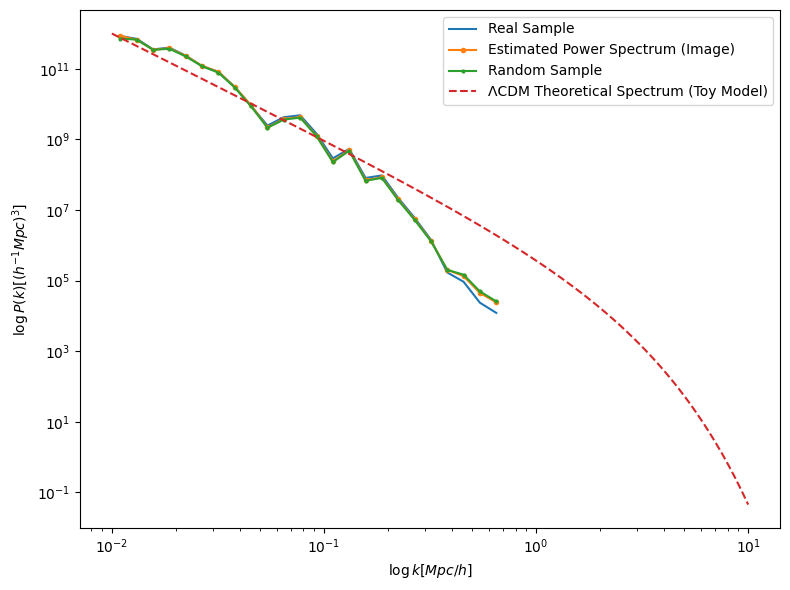}
    \caption{\footnotesize Post-processed matter power spectrum comparison between the original simulation data and a sample generated from the VAE’s latent space. A reference $\Lambda$CDM model, characterized by a $P(k) \propto k^{-3}$ behavior at small scales, is overlaid for comparison. In the intermediate range ($k \in [10^{-2}, 10^{-1}], h/\mathrm{Mpc}$), both the original and generated spectra show good agreement with the theoretical model, indicating that the VAE successfully captures the statistical structure of the training data. However, a noticeable suppression of power at high $k$ (small scales) is observed in the generated spectrum. This decay is likely due to the smoothing effects introduced by the VAE’s probabilistic generative process and the limited resolution of the input images.}
    \label{fig:5}
\end{figure}

Our results indicate that, on large scales, the generated power spectra closely match those of the training samples, demonstrating that the VAE successfully learns and reproduces the global statistical structure of the data. However, at lower scales, we observe minor discrepancies in the power spectra of the generated images compared to a simulation dataset (see figure \ref{fig:6} below). This deviation may be inherited from the original dataset itself, as imperfections and noise in training samples can propagate through the generative model. Despite these differences, the overall agreement in the power spectra suggests that the VAE captures the essential information required to replicate the structural features observed in cosmological simulations.

To further test the robustness of our results, we draw a random sample from the generated dataset and compute its power spectrum independently. The results remain fairly consistent with those obtained from the entire set, reinforcing the idea that the VAE correctly models the statistical properties of the data. These findings highlight the effectiveness of deep generative models in encoding and reproducing complex cosmological structures, providing a powerful tool for future applications in large-scale structure analysis and cosmic modeling.

\subsection{Non-parametric Power Spectrum Extrapolation}
\noindent
 
The power spectrum $P(k)$ recovered from the 2D Fourier transform is limited to scales of approximately $k \lesssim 0.3 h / \text{Mpc}$ due to the finite resolution of the input maps ($256\times 256$ pixels spanning a comoving size of $128 \text{Mpc}/h$). We employ Gaussian Process (GP) regression as a non-parametric method to extrapolate the power spectrum $P(k)$ beyond the range sampled directly from the simulated image data \cite{seikel2012reconstruction, zhang2018gaussian}. GPs offer several advantages over traditional parametric models when dealing with limited or noisy data, as is often the case in cosmological applications \cite{Velazquez:2024aya}. Unlike fixed functional forms such as power laws or exponential cut-offs, GPs model the underlying function as a distribution over possible functions, providing not only a best-fit prediction but also an estimate of the associated uncertainty. The smoothness and flexibility of the GP kernel allow us to capture subtle trends in the spectral data without overfitting, while the Bayesian nature of the method naturally quantifies extrapolation confidence through posterior variance.

The left panel of Figure \ref{fig:6} demonstrates this approach, where the blue line represents the regression fit by the GP, while the shaded around the fit line represents the confidence levels of $1\sigma$ and $2\sigma$ captured by the model. This is particularly useful in the small-scale regime ($k\gtrsim 1\;h/\text{Mpc}$), where nonlinear effects dominate and theoretical models become uncertain. By training the GP on the logarithm of
$P(k)$ and $k$, we ensure robustness across many orders of magnitude, maintaining consistency with the large-scale behavior while offering plausible predictions for smaller scales.

\subsection{Parametric Power Spectrum Extrapolation}

To investigate the behavior of the matter power spectrum at smaller scales and to allow comparisons with theoretical predictions, such as those of the $\Lambda$CDM model or HALOFIT \cite{2003MNRAS.341.1311S, takahashi2012revising}, it is necessary to extrapolate $P(k)$ toward higher wavenumbers, ideally reaching the nonlinear regime around $k \sim 1-10 \; h/\text{Mpc}$. For this purpose, we adopted a parametric approach, fitting a smooth empirical function to the portion of the spectrum that lies within a well-sampled and reliable intermediate range, specifically $10^{-3}<k<0.025 \; h/\text{Mpc}$. This interval was selected to avoid contamination from image resolution limits at high $k$ and noise-dominated behavior at low $k$. The functional form chosen for the fit captures the overall shape of the spectrum while remaining flexible enough to accommodate deviations from a pure power law. The adopted model is \cite{orjuela2024machine, prada2016hunting, alonso2024measurement}:

\begin{equation}
    P(k) = Ak^{-n}\cdot e^{-\alpha k} \cdot (1 + \beta k)^{-\gamma}.
\end{equation}
Here $A, \alpha, \beta \; \text{and} \; \gamma$ are the free parameters determined by nonlinear least squares fitting. This flexible model captures the expected power-law behavior on large scales and allows for a smooth suppression on small scales. The exponential term provides damping at high $k$, while the correction factor enables finer control of the curvature.

The calibrated model was then extrapolated to 
$k=10\;h/\text{Mpc}$, allowing comparison with theoretical spectra of the form $P(k) \propto k^{-3}$. This procedure enables a continuous, physically-motivated extension of the estimated power spectrum, bridging the gap between the image-derived data and cosmological theory at non-linear scales.

\begin{figure}[t!]
    \centering
\begin{minipage}[b]{0.49\textwidth}
    \includegraphics[height=5.2cm, width=7.8cm]{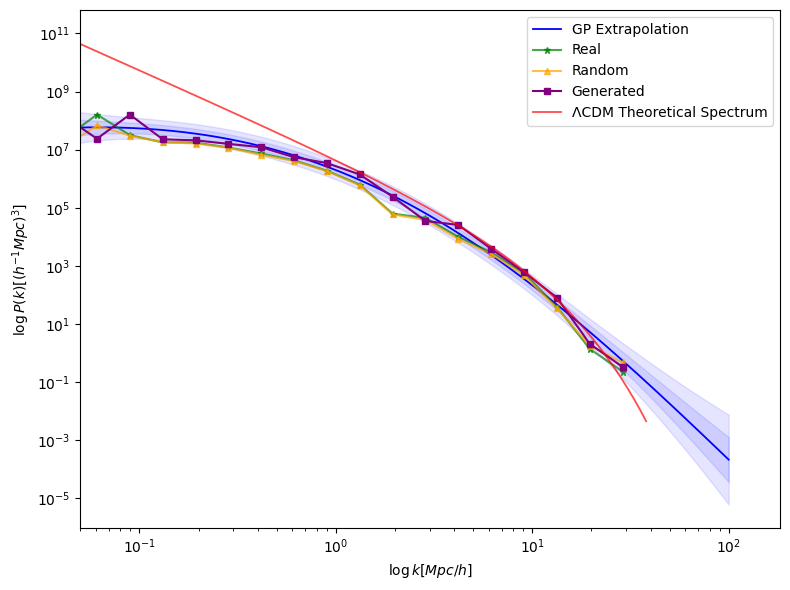}
 \end{minipage}
  \begin{minipage}[b]{0.49\textwidth}
    \includegraphics[height=5.2cm, width=7.8cm]{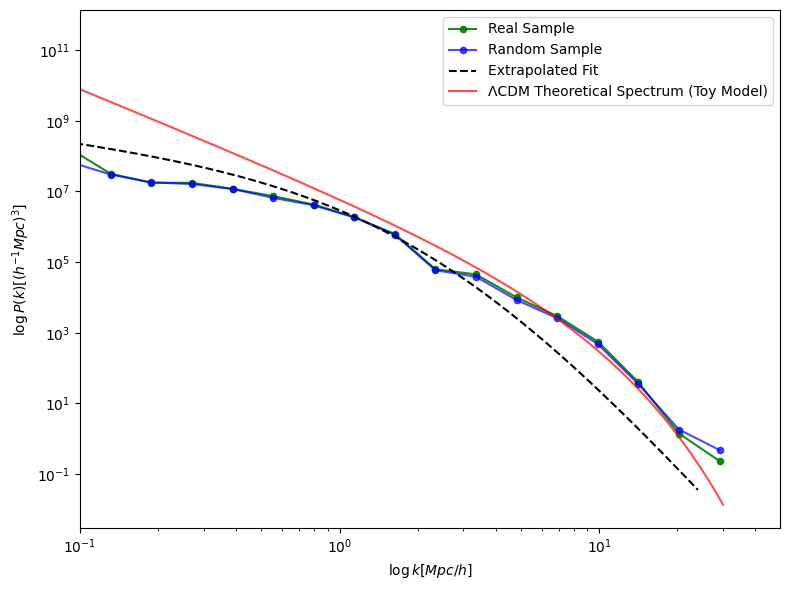}
  \end{minipage}
  \caption{\footnotesize The extrapolated power spectrum, derived from both the training and VAE-generated images, is compared with the theoretical prediction from the $\Lambda$CDM model. \textit{Left:} Gaussian process regression. \textit{Right:} Parametric extrapolation with damping function. At intermediate scales ($k \sim 10^{-1}$ to $k \sim 10^{1} , h/\mathrm{Mpc}$), the spectra exhibit excellent agreement, capturing the expected shape and amplitude of the matter distribution. This consistency suggests that the VAE successfully encodes and reproduces the statistical features present in the training data. However, at larger scales (low $k$), a discrepancy emerges between the theoretical model and the generated spectra. While the overall dark matter content at these scales remains similar—as confirmed after a scale adjustment during postprocessing—the VAE-generated spectrum shows a slight suppression of power compared to $\Lambda$CDM. This deviation is likely attributable to the inherent smoothing introduced by the VAE's probabilistic generative process, combined with the limited resolution of the input images used for training and generation.} \label{fig:6}
\end{figure}

The right panel of Figure \ref{fig:6} shows the model used for this extrapolation. This particular model arises from previous studies of the transfer function models, such as the Dicus transfer function, which, applied to the context within the Zel'dovich approximation \cite{1970A&A.....5...84Z, white2014mock}, also indicates that for strictly cold dark matter, the power spectrum exhibits an asymptotic decay proportional to $k^{-3}$
at small scales, as discussed in Konrad et al. (2022) \cite{konrad2022asymptotic}. These and other findings \cite{2009MNRAS.397.1275W,ginat2025gravitational} support the adoption of a $P(k)\propto k^{-3}$ extrapolation in the non-linear regime. Therefore, when extending our power spectrum beyond the range directly accessible through the Fourier transform of images, we employ a model with a slope of $-3$ to adequately represent the expected small-scale behavior.

\subsection{{Power Spectrum Ratios and Cross-Coherence Validation}}
  
In order to assess the fidelity of the VAE-generated cosmological fields, we perform a Fourier-space comparison between the simulated, generated, and random density maps. We evaluate the 1D matter power spectrum $P(k)$, the ratio $P_{VAE}(k)/P_{sim}(k)$, and the cross-spectral coherence between the generated and real fields. These metrics provide complementary information: the amplitude ratio quantifies how well the VAE reproduces the two-point statistics of the underlying density field, while the coherence captures phase agreement, and thus tests the preservation of spatial structure.

%Together, these diagnostics allow us to identify the scales where the VAE is faithful to the physical signal and where deviations arise due to resolution limits, noise, or incomplete representation of small or large scale modes.

\begin{figure}[t]
    \centering
    \includegraphics[width=0.8\linewidth, height =0.75\linewidth]{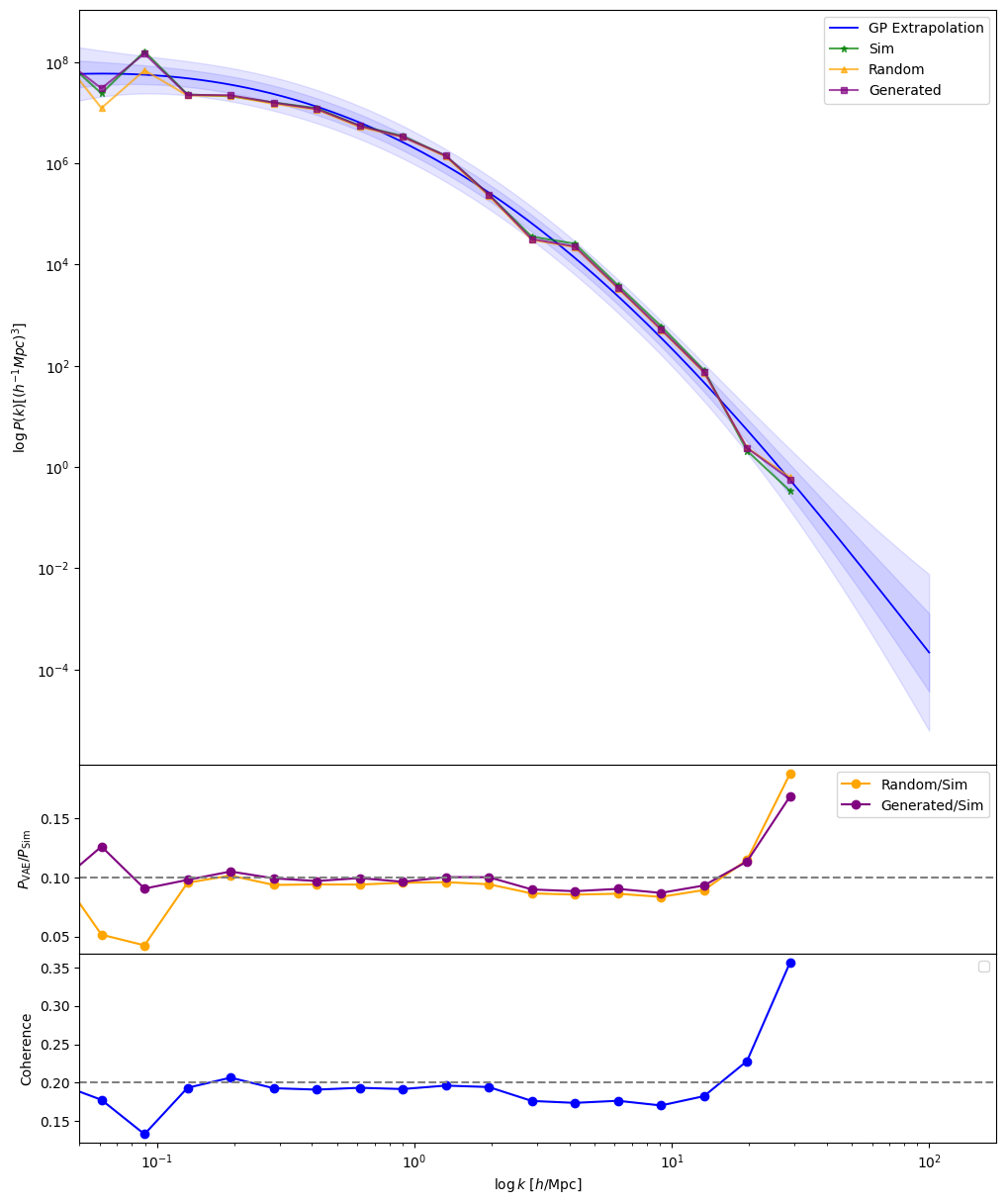}
    \caption{\footnotesize {Fourier-space comparison between real, generated, and random dark-matter density fields. \textit{Top}: The 1D matter power spectra $P(k)$ extracted from the real simulation (green), a VAE-generated sample (purple), and a random control field (orange), compared against a Gaussian-process extrapolation (blue). The shaded bands indicate the 1$\sigma$" and 2$\sigma$ uncertainties of the GP model. The VAE spectrum closely follows the simulated $P(k)$ over the intermediate range $0.1 \leq k \leq 5\;h /\rm Mpc$, where numerical artifacts are minimal and physical modes are well sampled. \textit{Middle}: Ratio of the generated and random spectra with respect to the real simulation. The VAE-to-simulation ratio remains near unity across the mid-range of scales, indicating that the model accurately captures the two-point statistics of the field where structure formation is most prominent. Deviations at low $k$ are driven by finite-box effects and poor sampling of long-wavelength modes, while high-$k$ discrepancies reflect resolution limits and VAE-induced smoothing. \textit{Bottom}: Cross-spectral coherence between the VAE-generated and real maps. Coherence is high across the same intermediate band of wavenumbers, showing that the VAE preserves phase information and thus reproduces the spatial arrangement of filaments and voids. The drop in coherence at low and high $k$ reflects the lack of well defined super modes and suppressed small scale fluctuations, respectively.}}
    \label{fig:7}
\end{figure}

In figure \ref{fig:7}, we observe that the comparison between simulated, VAE–generated, and random fields reveals that the agreement between the power spectra is strongest across the intermediate range of scales, approximately $0.1 \leq k \leq 5\; h/\rm Mpc$. In this regime, the ratio $P_{VAE}(k)/P_{sim}(k)$ remains close to unity, indicating that the VAE successfully captures the dominant two-point statistical properties of the matter distribution. At low wavenumbers (large scales), $k \leq 0.1\; h /\rm Mpc$, deviations are expected and arise primarily from the limited number of long-wavelength modes available within the finite simulation box; this affects both the input data and the VAE reconstructions in the same manner. In contrast, high-k (small scales) discrepancies reflect the combined effects of the pixel resolution of the projected density maps and the intrinsic smoothing tendency of convolutional generative models, which damp small scale fluctuations beyond the Nyquist frequency of the input. These behaviors define the natural scale boundaries within which the VAE can be reliably evaluated using Fourier-space statistics.

Additionally, it is essential to verify whether the VAE also preserves the phase information that encodes the spatial arrangement of cosmic structures. To assess this, we compute the cross-spectral coherence between the generated and simulated fields. In figure \ref{fig:7}, we can observe that coherence remains high across the same intermediate range of scales where the power-spectrum ratio is close to unity, indicating that the VAE not only reproduces the correct clustering strength but also maintains the relative positioning of filaments and voids in Fourier space. At low (high) $k$, the coherence decreases (increases), reflecting the limited number of long-wavelength modes and the smoothing of the small scale structure, respectively. This complementary diagnostic confirms that the VAE retains physically meaningful spatial correlations on the scales where the underlying signal is robust.

\subsection{{Probability density functions and moment analysis}}

\begin{figure}
  \begin{minipage}[b]{.45\linewidth}
    \centering
    \includegraphics[width=1\linewidth]{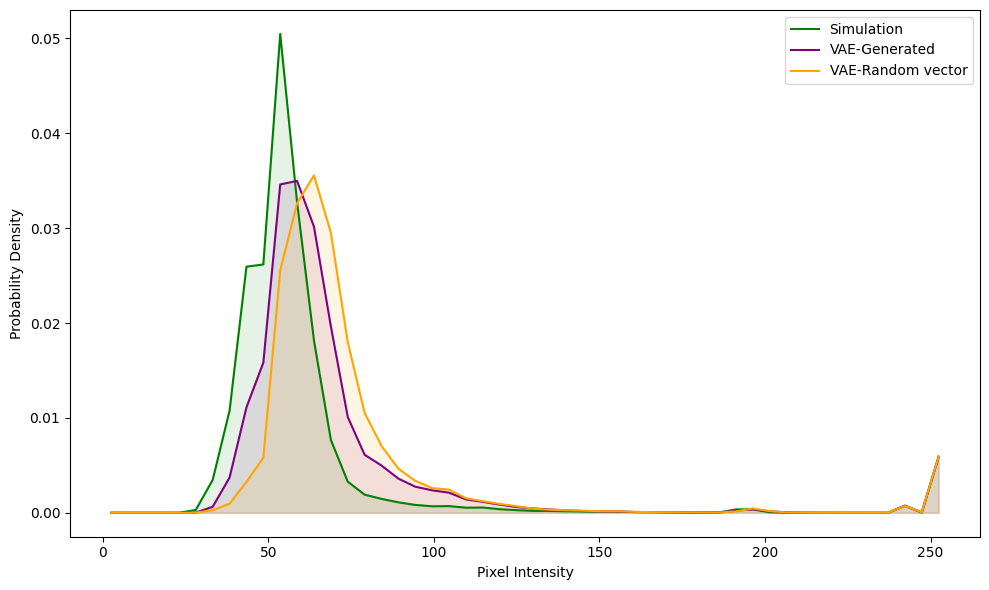}
    \captionof{figure}{\footnotesize Probability density functions of pixel intensities for the simulation cosmological map (green), a VAE-generated map obtained from a real encoded latent vector (purple), and a VAE-generated map obtained from a random latent vector (orange). All three Probability Density Functions exhibit similar peaks, tails, skewness, and kurtosis, demonstrating that the VAE latent space captures the full one-point statistical structure of the cosmological field. The close overlap between the real and random-latent PDFs confirms that decoding random samples from the VAE prior produces new realizations consistent with the true underlying distribution.}\label{fig:8}
  \end{minipage}\hfill
  \begin{minipage}[b]{.45\linewidth}
    \centering
    \begin{tabular}{ |p{1.35cm}|p{.7cm}|p{.7cm}|p{1.2cm}|p{1cm}|  }
 \hline
 \multicolumn{5}{|c|}{Summary Statistics} \\
 \hline
 Sample & $\mu$ & $\sigma$ & Skewness & Kurtosis\\
 \hline
 Simulation   & 62.62 & 39.53 & 4.01 & 16.05\\
 Generated    & 71.61 & 39.03 & 3.64 & 13.82\\
 Random Vector& 75.95 & 37.91 & 3.69 & 14.15\\
 \hline
\end{tabular}
    \captionof{table}{Summary statistics of the selected image samples.}\label{table:2}
  \end{minipage}
\end{figure}

While the power spectrum and coherence characterize the spatial correlations of the density field, it is equally important to validate the one-point distribution of pixel intensities, which correspond to projected matter densities in our dataset. We therefore compute the normalized probability density function (PDF) of pixel values for the real images, the VAE-generated images, and the random field, as shown in figure \ref{fig:8}. The probability distribution functions of the real and generated images show a high level of agreement: both exhibit a sharp peak near the mean density, an extended tail toward higher intensities, and a rapid falloff for low values. In contrast, the random control displays a broader and less physically motivated distribution, as expected.

To quantify these trends, we compute the skewness and kurtosis of each distribution, as shown in table \ref{table:2}. The real image shows an indicative of the strong non-Gaussian tail produced by nonlinear gravitational collapse. The VAE-generated images yield a similar skewness for both random and generated samples, demonstrating that the model successfully captures the asymmetry of the density field. The kurtosis values for the real sample and the VAE outputs also indicate that the generator reproduces the heavy-tailed behavior characteristic of cosmic density distributions.

The similarity between the real, generated, and random-latent images across all statistical metrics, including the pixel intensity via the Probability Density Function, skewness and kurtosis, and the shape of the power spectrum, demonstrates that the VAE has successfully learned a latent space that captures the key statistical properties of the cosmological density field. Because the encoder regularizes the latent space toward a standard multivariate Gaussian, random vectors drawn from this prior lie on the same learned manifold as the encoded real images. Thus, the random latent samples effectively represent new cosmological realizations drawn from the generative model.

\subsection{{Latent-Space Diagnostics and Representation Analysis}}

To investigate latent space encoding, we generated 24 latent vectors associated with the same physical configuration plus random variations over vector values, and created projections on a 2D manifold. The resulting distribution shows no evidence of disjoint clusters or separated substructures, but instead forms a compact and continuous region of latent space. This behavior is consistent with the regularization proceeding directly from the model, which promotes an approximately Gaussian prior and ensures that nearby points decode into statistically similar density fields.

\begin{figure}[b]
    \centering
\begin{minipage}[c]{0.49\textwidth}
    \includegraphics[height=5.2cm, width=7.8cm]{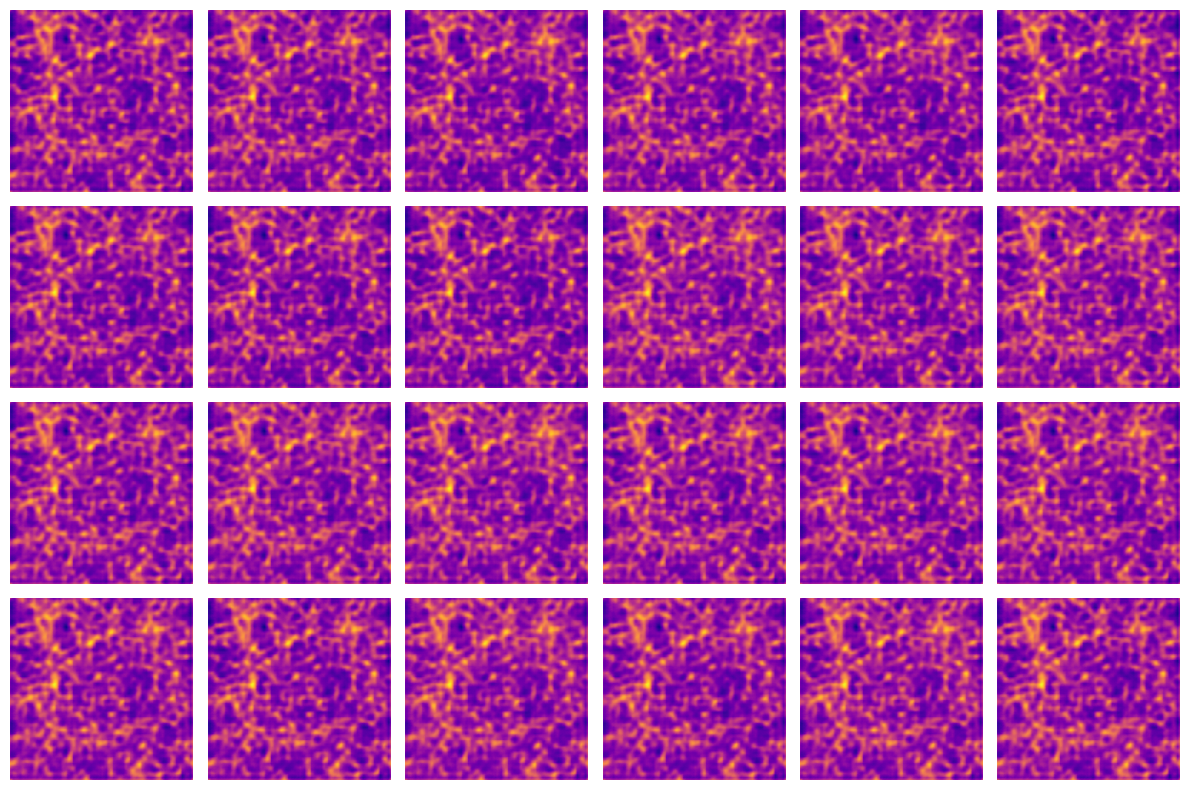}
 \end{minipage}
  \begin{minipage}[c]{0.49\textwidth}
    \includegraphics[height=5.2cm, width=7.8cm]{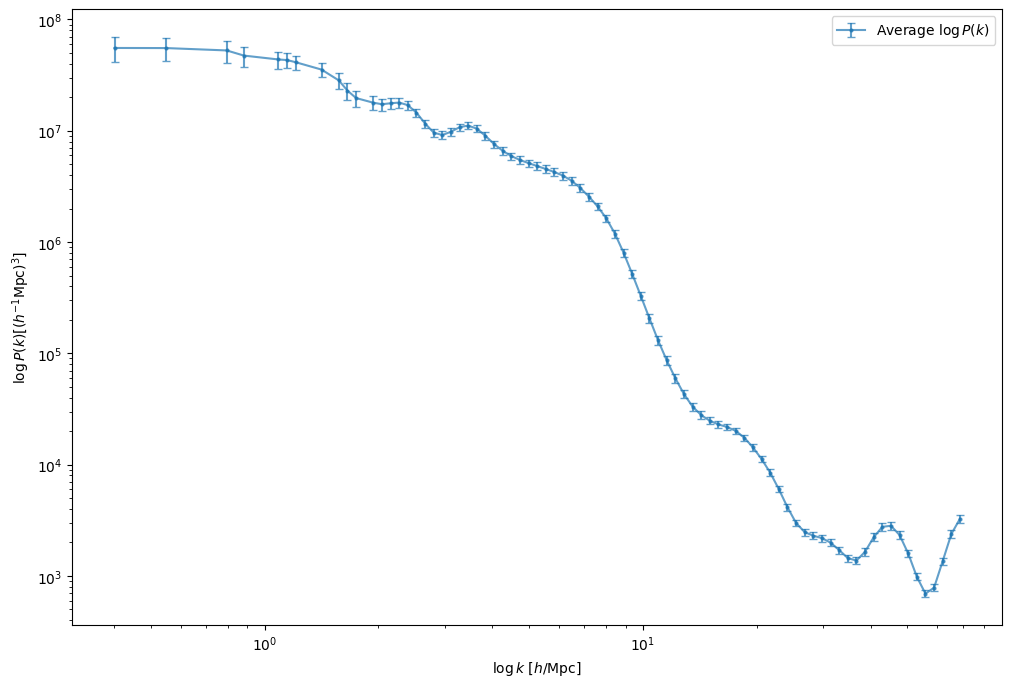}
  \end{minipage}
  \caption{\footnotesize \textit{Left:} 2D projection of the VAE latent vectors, showing a compact and continuous manifold with no artificial clustering. \textbf{Right:} Mean matter power spectrum of the corresponding decoded images with 2$\sigma$ error bars. The small variance across most scales indicates that latent variations produce statistically consistent cosmological realizations.} \label{fig:9}
\end{figure}

We then decoded each latent vector into an image and computed its matter power spectrum. The mean spectrum, shown in figure \ref{fig:9}, closely matches the expected nonlinear shape, with small dispersion across most wavenumbers. This indicates that the latent space encodes stable and physically meaningful variations of the density field and that different latent samples represent valid realizations of the same cosmological structure rather than random noise. The larger scatter observed at high 
$k$ reflects sensitivity to pixel variations and does not affect the nonlinear regime. These results demonstrate that the latent space is organized in a smooth and physically coherent manner, faithfully capturing the underlying statistical properties of the cosmological field.

\begin{figure}
    \centering
    \includegraphics[width=0.7\linewidth]{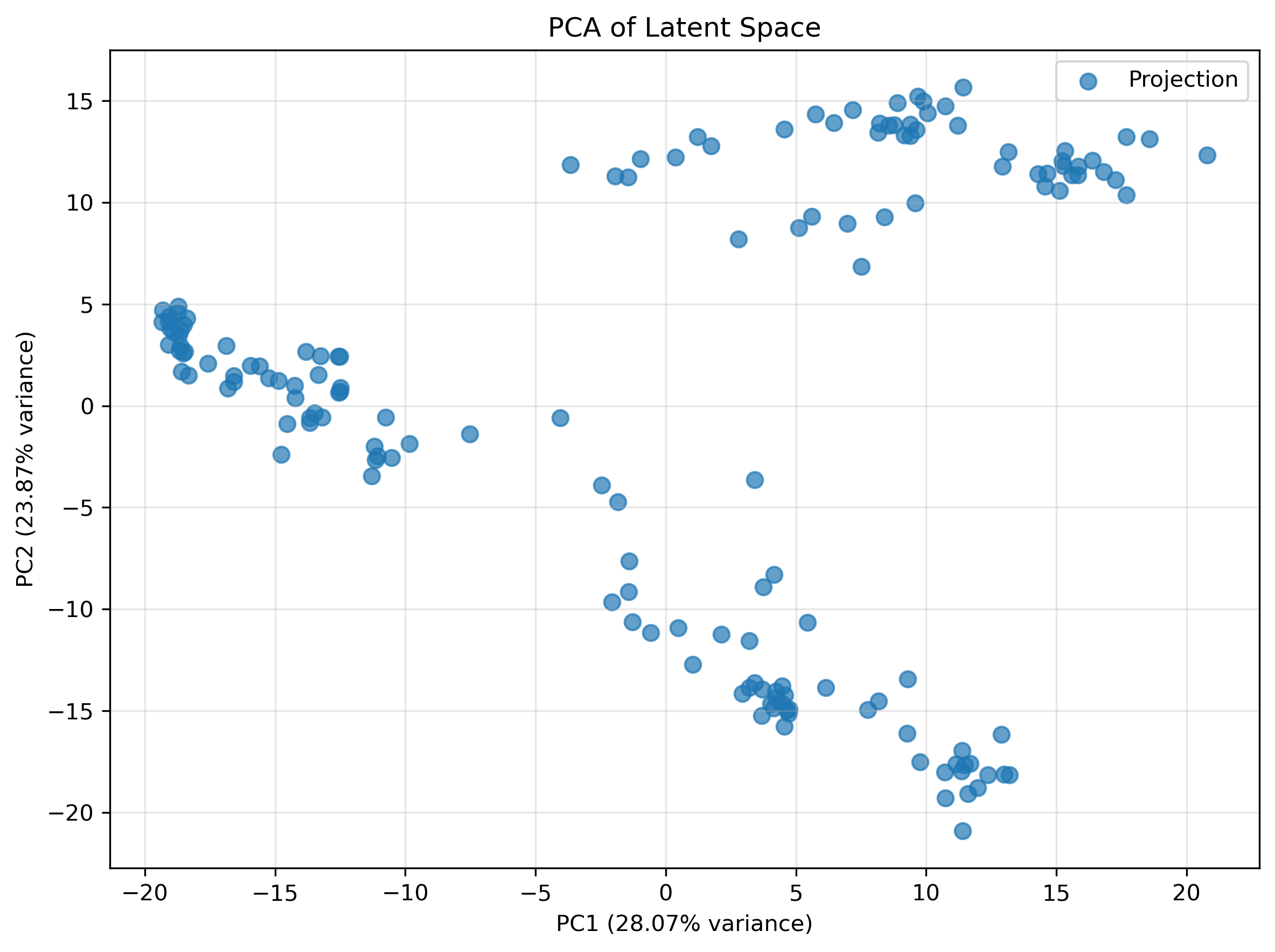}
    \caption{\footnotesize PCA projection of the encoder latent means for projected dark-matter density slices. The first two components capture the dominant variance directions in the latent space and provide a qualitative diagnostic of its structure.}
    \label{fig:latent_pca}
\end{figure}

To provide qualitative insight into the learned representation, we project the encoder latent means $\mu(x)$ of projected density slices into two dimensions using PCA. In Fig.~\ref{fig:latent_pca}, the first two principal components capture $\sim 50\%$ of the variance in this 2D visualization. The projection shows a structured but continuous distribution, as expected when compressing a high-dimensional latent space to two components; this indicates that the dominant latent variations are organized along a few leading directions rather than being random.

%%%----------------------------------------------------%%
\section{Final discussion and future work}\label{sect:5}
%%%----------------------------------------------------%%
\label{sec:conclusions}

In this work, we demonstrated the feasibility and effectiveness of using a variational autoencoder to compress, reconstruct, and generate synthetic cosmological images derived from dark matter simulations. Visual inspections confirm that the VAE captures key morphological features, such as filaments and halos, suggesting a successful encoding of the large-scale structure into the latent space. Although some fine-scale details are smoothed out, which is a common expected limitation of probabilistic generative models, the generated samples remain statistically representative of the training set. This is evidenced by the diversity of generated outputs and their consistency with known cosmological structures, indicating that the model generalizes well rather than memorizing the training data.

Quantitative analysis of the generated images through power spectrum measurements shows good agreement with the expected $\Lambda$CDM model, particularly at intermediate scales. Small-scale discrepancies, attributed to resolution limitations and the inherent smoothing of the VAE, were mitigated through filtering and empirical extrapolation using a physically motivated fitting function. The extrapolated spectra recover the expected $k^{-3}$ behavior in the non-linear regime, consistent with theoretical predictions.

These results underscore the potential of VAEs not only as tools for compression and efficient reconstruction but also as generative models capable of producing statistically coherent, physically meaningful dark matter density fields. This generative capability enables the rapid production of synthetic datasets, opening the door to scalable data augmentation strategies that can support the training of more sophisticated models or serve as input for simulation-based inference tasks. This approach represents a computationally efficient and versatile approach for modeling, generating, and analyzing large-scale structures in cosmology.

Future work will focus on extending this approach to higher-resolution simulations and larger image sizes to better capture small-scale structures and improve the fidelity of the recovered power spectra. Increasing the resolution and expanding the comoving volume would enable access to a broader range of physical scales and more accurate modeling of nonlinear clustering. Additionally, incorporating Physics-Informed Neural Networks\cite{RAISSI2019686, 2021NatRP...3..422K} into the generative process offers a promising field to improve physical consistency by embedding cosmological equations (such as dynamics and thermodynamics) directly into the learning framework. Recent advances in this area demonstrate that the combination of deep generative models with physical constraints can improve both accuracy and interpretability \cite{MODI2021100505, zhu2019physics, taufik2025latentpinns}. Integrating such techniques with variational autoencoders or with more advanced architectures, such as diffusion models, may enable the generation of synthetic cosmological data that resembles theoretical expectations more closely, bridging the gap between deep generative learning and high-precision cosmology.

\section*{Acknowledgments}
%%%---------------------------------------------------------%%
\noindent
J.Ch. thanks SECIHTI for providing scholarship support during the post-graduate research and the support of ICF-UNAM and CIC-IPN. J.A.V. acknowledges support from FOSEC SEP-CONACYT Ciencia B\'asica A1-S-21925,  UNAM-DGAPA-PAPIIT IN117723, IN110325 and Cátedra de Investigación Marcos Moshinsky. 
%%%---------------------------------------------------------%%

%\newpage
% \bibliographystyle{abbrv}
\bibliographystyle{unsrt}
\bibliography{main.bib}

\end{document}